\documentclass{INTERSPEECH2023}
\usepackage{todonotes}
\usepackage{array,multirow,graphicx}
\usepackage{svg}

\newcolumntype{L}[1]{>{\raggedright\arraybackslash}p{#1}}

\newcolumntype{C}[1]{>{\centering\arraybackslash}p{#1}}

\newcolumntype{R}[1]{>{\raggedleft\arraybackslash}p{#1}}

\interspeechcameraready

\title{An Integrated Algorithm for Robust and Imperceptible \\Audio Adversarial Examples}
\name{
Armin Ettenhofer$^{1,2}$,
Jan-Philipp Schulze$^{2}$,
Karla Pizzi$^{1,2}$
}
\address{
  $^1$Fraunhofer Institute for Applied and Integrated Security (AISEC), Germany \\
  $^2$TUM School of Computation, Information and Technology, Technical University Munich, Germany
  }
\email{\{armin.ettenhofer,
jan-philipp.schulze,
karla.pizzi\}@tum.de
}

\begin{document}

\maketitle
\begin{abstract}
Audio adversarial examples are audio files that have been manipulated to fool an automatic speech recognition (ASR) system, while still sounding benign to a human listener.
Most methods to generate such samples are based on a two-step algorithm: first, a viable adversarial audio file is produced, then, this is fine-tuned with respect to perceptibility and robustness.
In this work, we present an integrated algorithm that uses psychoacoustic models and room impulse responses (RIR) in the generation step.
The RIRs are dynamically created by a neural network during the generation process to simulate a physical environment to harden our examples against transformations experienced in over-the-air attacks. 
We compare the different approaches in three experiments: in a simulated environment and in a realistic over-the-air scenario to evaluate the robustness, and in a human study to evaluate the perceptibility.
Our algorithms considering psychoacoustics only or in addition to the robustness show an improvement in the signal-to-noise ratio (SNR) as well as in the human perception study, at the cost of an increased word error rate (WER).
\end{abstract}
\noindent\textbf{Index Terms}: adversarial examples, automatic speech recognition, psychoacoustic masking, robustness

\section{Introduction}
Neural networks (NNs), the technology at the heart of automatic speech recognition (ASR) systems, have an inherent weakness called adversarial examples \cite{abdullah2021sok}. 
These are manipulated inputs that fool an ASR system, while seeming benign to a human listener.

For audio adversarial examples to be successful in an over-the-air scenario, the added attack patterns must withstand real-world physical phenomena, e.g., reflections and overlays, up to the point they are re-recorded by a microphone.
While conveying the attack message to the ASR, these added sounds should be imperceptible to humans.
We describe such changes by the room impulse response (RIR), an approximation for the transformations that the audio experiences during playback.
In our work, we randomly sample RIRs to make our audio adversarial examples robust against physical phenomena previously impairing the attack success.
Imperceptibility can be optimized using a psychoacoustic metric, which models the perception of the human auditory system. 

Carlini~and~Wagner~\cite{carlini2018audio} were among the first to investigate targeted adversarial examples in the audio domain, managing to create adversarial perturbations that can trick NN-based ASR systems. 
Their method, the CW approach, worked as a proof of concept and pointed out future challenges for attacking ASR, i.e., making the perturbation imperceptible and robust against over-the-air transmission. 
Qin~et~al.~\cite{qin2019imperceptible} expanded on this work by developing methods to generate robust and imperceptible adversarial examples. 
To achieve robustness against transformations, they used the expectation over transformation \cite{athalye2018synthesizing} (EOT) method previously successfully used for adversarial examples in the visual domain. 
In order to make the perturbation imperceptible, Qin~et~al.\ applied psychoacoustics to estimate the perceptibility of the perturbation.
This advanced approach models human audio perception better than a simple distance metric.
However, they only evaluated the robustness of their adversarial examples in a simulated environment. 
Psychoacoustic properties were also analyzed by Schönherr~et~al.~\cite{schonherr2020imperio}, who developed robust adversarial examples that work in over-the-air conditions and conducted detailed evaluations for their specific DNN-HMM system under attack.
Their attack makes use of a convolutional layer to simulate the audio's transmission through the room.
Dörr~et~al.~\cite{dorr2020towards} recognized another challenge with over-the-air attacks: 
because of the Mel-frequency cepstral coefficients (MFCC) feature extraction that many ASR models use during pre-processing, there is a risk of overfitting to a specific window alignment.
The MFCC extraction depends on the temporal offset of the signal, thus the window alignment in a physical setup will never be precisely the same. 
To prevent this overfitting, they applied random temporal offsets during training. 
Another example of physical attacks using adversarial examples is the work of Yakura~and~Sakura~\cite{ijcai2019p741}, who adapted the EOT algorithm to generate robust adversarial examples using a band pass filter, Gaussian noise, and RIRs. 
However, while successfully tricking the model, they could only generate adversarial examples with very high signal-to-noise ratios and only for 2-3 word sentences. 
Chen~and~Sakuma~\cite{chen2020metamorph} reported successful over-the-air adversarial examples over distances up to 6m, even without a line of sight between speaker and listener. 
In addition to EOT, they also used domain discrimination training to remove bias from the measuring sources of their RIRs. 
Additionally, they tried to make their examples inconspicuous by shaping them like natural background noise, e.g., bird chirps.

Although much work was done on developing different adversarial examples, the results are hard to compare because they rarely use the same model or experimental setup. 
This problem is addressed by Olivier~and~Raj~\cite{olivier2022recent}, who developed an extensible framework, called robust\_speech.
It incorporate implementations for many audio adversarial attacks, both targeted and untargeted, and utilities for training, attacking, and evaluating models, providing an overview of current models, attacks, and defenses. 
The framework is based on SpeechBrain \cite{speechbrain}.

In this work, we build upon robust\_speech, while learning from the latest advances on physical adversarial examples in computer vision \cite{chen2019shapeshifter,10.1145/3560830.3563733}, thus improving physical audio adversarial examples.
We expand upon previous methods, particularly that of \cite{qin2019imperceptible}, and suggest several adaptions: 
(1) We implement and evaluate the method of dynamically generating the RIRs used for training robust examples, which could prevent overfitting to biases in the used RIR dataset and provide more flexibility. 
(2) We expand upon the psychoacoustic model used by \cite{qin2019imperceptible} and implement a more detailed approximation, to make generated adversarial examples less perceptible.
(3) We include all these proposed methods into one integrated algorithm and compare this to the CW approach \cite{carlini2018audio}.
In our evaluation, we use a simulated environment, real over-the-air tests, and a human study.

\section{Method}
Audio adversarial examples are inputs to automatic speech recognition systems, specifically generated to cause a false transcription $t$, while sounding benign to the human listener.
Typically, these instances can be defined as minimizing the adversarial loss $\mathcal{L}$ that calculates the distance between the adversarial example's classification $f\left(x+\delta\right)$ and the target $t$ using the model's loss function $l$.
While the adversarial noise $\delta$ is limited by some parameter $\epsilon$, the loss takes some norm into account to control that the adversarial noise remains small. The impact of the regularization term is controlled via $\alpha$:
\begin{align*}
    \min \mathcal{L} = \, &l \left(f \left(x + \overline{\delta} \right), \, t \right) + \alpha \Vert \delta \Vert \\
    \text{such that }& \delta \leq \epsilon.
\end{align*}
This optimization problem is usually solved using iterative gradient descent.

In this work, we improve upon existing audio adversarial attacks, creating audio files that work in an over-the-air scenario and are imperceptible or at least inconspicuous to human listeners. 
We incorporate room impulse responses and psychoacoustic masking into our integrated approach in which we replace a commonly two-step algorithm by just one optimization loop with one loss function, see Section~\ref{ssec:IntegratedApproach}.

\subsection{Psychoacoustic masking}\label{ssec:Psychoacoustic}
The first implementations for audio adversarial examples approximated the human hearing with norms taken in the time domain, i.e., $\text{dB}(x)$. 
The mechanical foundations that make up hearing and perception and their function are, however, better analyzed in the frequency domain.
This enables us to account for auditory masking.
Auditory masking is a phenomenon by which loud impulses at a frequency make smaller impulses at certain other frequencies inaudible, i.e., mask them. 
These mechanisms are, e.g., used in compression \cite{iso1993information} to discard sounds that are imperceptible to the listener.
This phenomenon can also be used to hide adversarial changes made to the audio. 
Thus, we can use it to hide the manipulations in our adversarial examples from human listeners. 

\begin{figure}
    \centering
    \includegraphics[width=0.45\textwidth]{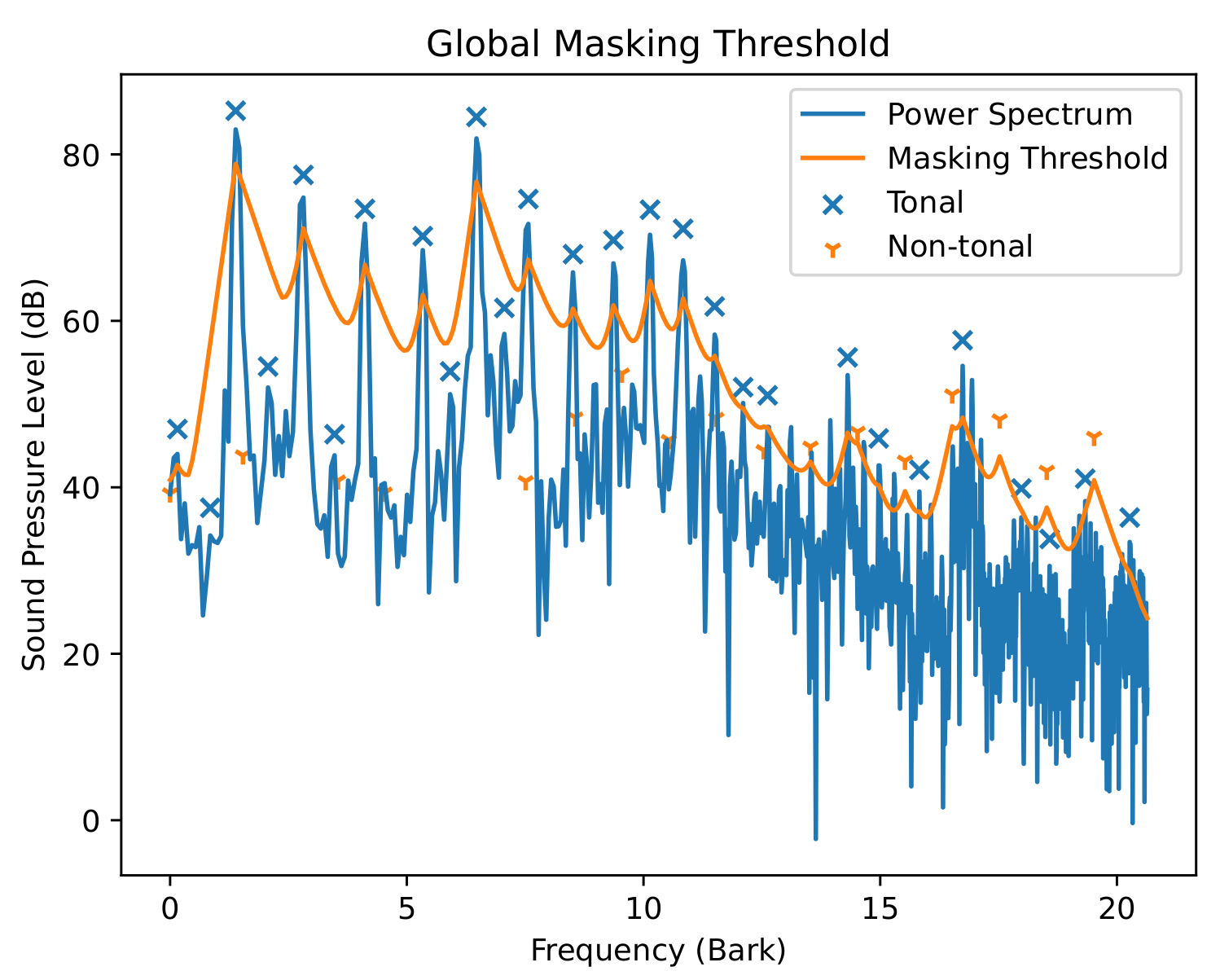}
    \caption{
    The psychoacoustic model considers the human audio perception.
    At certain frequencies, small impulses are inaudible.
    We show the masking threshold in orange.
    All sounds below this line are considered to be masked.}
    \label{fig:threshold}
\end{figure}

\begin{figure*}[h]
    \centering
    \includegraphics[width=0.8\textwidth]{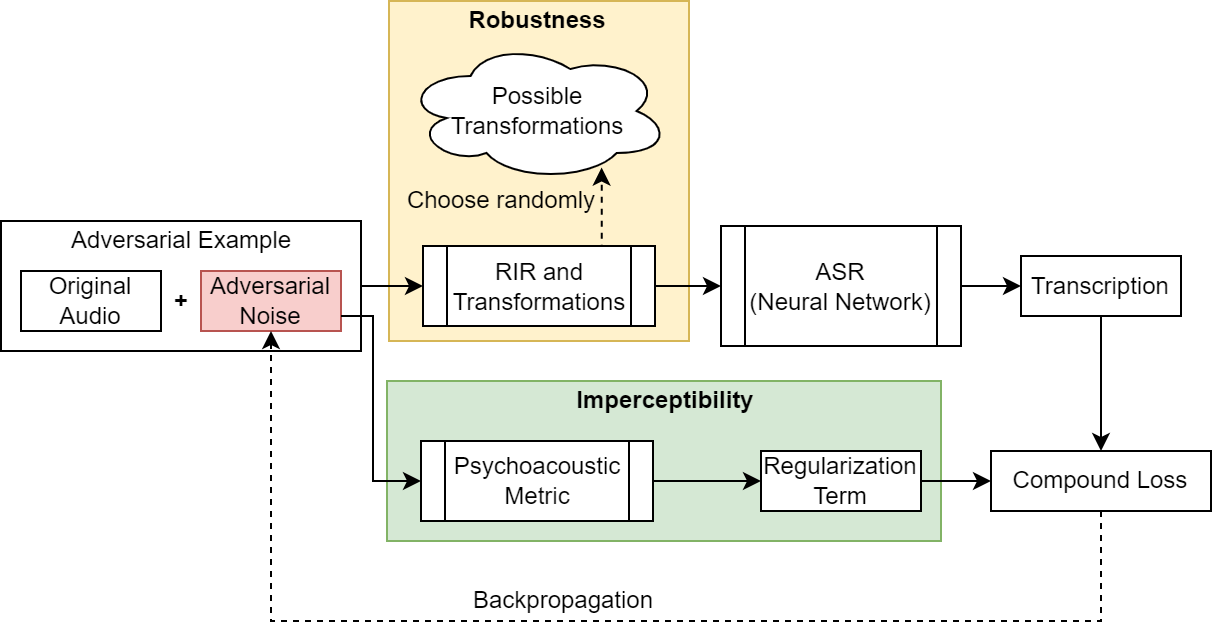}
    \caption{Integrated algorithm: psychoacoustics and robustness are both considered in one single compound loss.
    There is only one single optimization step to be repeated iteratively.}
    \label{fig:OneStepAlgorithm}
\end{figure*}

For optimizing the imperceptibility of our adversarial noise, we use the psychoacoustic model from the MPEG audio standard \cite{lin2015audio, iso1993information}.
It approximates perceptibility based on the mechanism of the human auditory system using the concept of auditory masking.
The psychoacoustic model was also used in a simplified form by \cite{qin2019imperceptible} and \cite{schoenherr-19-psychoacoustic} in their attack. 
However, \cite{qin2019imperceptible} classifies every masker as a tonal masker, while we differentiate tonal and noise-like maskers per the original model, which have different effects on masking, see Figure~\ref{fig:threshold}.
Similar to both, we pre-compute the psychoacoustic masking, as this remains constant during all iterations.

We use the implementation specification from \cite{lin2015audio} to build a global simultaneous masking threshold that takes into account masking effects that appear while the masker is present.
The global simultaneous masking threshold is achieved by combining the individual simultaneous masking threshold and absolute threshold of hearing.
Then, we calculate the sound pressure level of the perturbation during every iteration and take the mean of any values above the masking threshold as our regularization term, see Figure~\ref{fig:OneStepAlgorithm}.
Figure~\ref{fig:threshold} displays an example masking threshold (in orange): all sounds below this line are masked and would thus not be perceived by a human listener.

\subsection{Room impulse responses}\label{ssec:RoomImpulse}
In order to use adversarial examples in the real world, they should work via the so-called air gap.
For images, this means that they should work even, e.g., when printed and detected by a camera.
For audio files, this means they should work even when played via a speaker and recorded by a microphone.
For this, Athalye et al.~\cite{athalye2018synthesizing} developed the so-called expectation over transformation (EOT) algorithm to make adversarial examples robust against any transformations:
during training, at every step of the gradient descent, a random transformation similar to actual transformations in the physical world is applied to the adversarial example before passing it to the model. 
For each iteration, one draws a new transformation. 

To make our adversarial examples robust against three kinds of transformations that occur in over-the-air attacks, we implement an adaptation of the EOT algorithm.
First, in every iteration, we add a small amount of Gaussian noise to the input to simulate background sounds that occur in real scenarios. 
Next, we use room impulse responses (RIR), mathematical expressions of the transformations that happen to audio transformations experienced by the audio when the sound waves bounce around the room. 
We apply them to a signal to simulate how the signal would be transformed if played and recorded in the room.
We create eight copies of the adversarial example and convolve them with different RIRs. 
Rather than choosing the RIR from a potentially small pool of measured or pre-generated RIRs, we generate them dynamically, i.e., a completely new RIR is chosen for every iteration.
This may prevent overfitting to certain RIRs or a bias in the RIR pool.
These RIRs are generated using the FAST-RIR model\footnote{See \url{https://github.com/anton-jeran/FAST-RIR}.} \cite{9747846}, a NN that takes the dimensions and sound absorption of the room along with the position of the listener and speaker as input and returns the predicted RIR. 
Importantly, it is fast enough to be used during optimization. 
The inputs to the FAST-RIR model are chosen from a configurable range of rooms and calculated using batch processing. 
 
Finally, to each of the eight resulting versions of the adversarial example, we add a random offset to the start of the audio.
The goal of the offset is to simulate the misalignment that happens during playback and recording:
As a preprocessing step to the ASR, MFCC binning is performed, which converts the signal into a frequency space representation with multiple frames. 
As the first step of this is a short-term Fourier transform, the optimization might overfit to a particular window alignment. 
As shown by \cite{dorr2020towards}, using a random offset prevents this.
The offset is implemented by shifting the values in each vector, filling with zeroes. 
To select the best adversarial example, we check for the adversarial examples that were successful for most of the eight room impulse responses.
Out of these, we take the one with the least perceptibility.

\subsection{Integrated approach}\label{ssec:IntegratedApproach}
We adapt the two-step method presented by \cite{qin2019imperceptible} with a single-step optimization process.
While \cite{qin2019imperceptible} first produces a viable adversarial audio file and then fine-tunes it with respect to perceptibility and robustness, we propose to use an integrated approach that immediately prioritizes both. 

Our integrated approach takes the original audio file and the target phrase as input and returns an {\color{gray}audio adversarial example} as output. 
Initially, we start with the original audio file $x$.
Then, a perturbation $\delta_0$ is added to the original audio and the loss is calculated, see Figure~\ref{fig:OneStepAlgorithm}. 
Based on this, we perform backpropagation to calculate the gradient of the input in relation to the loss.
With this, we update the perturbation $\delta_i$. 
We repeat these steps until we have found a perturbation that leads to the correct target phrase $t$. 
In order to account for {\color{red}robustness} and {\color{blue}imperceptibility}, instead of only optimizing on the model loss $l$, the optimizer uses a compound loss $\mathcal{L}$: 
\begin{align*}
    \min \limits_{\delta} \mathcal{L} = \, &l \left(f \left( {\color{red}0 \Vert} \left(\left({\color{gray}x + \overline{\delta}} {\color{red}+ g} \right) {\color{red}* r} \right)\right), \, t \right) {\color{blue}+ \alpha \cdot \beta \cdot p\left( \delta \right)} \\
    \text{such that }& \overline{\delta}_i = \min \left( \delta_i, \, \epsilon \right) \\
    & \beta = \sqrt{\frac{\text{len}(x_{\text{ref}})}{\text{len}(x)}}.
\end{align*}
Here, $l$ refers to the model loss measuring the distance of the transcription $f$ to the target phrase $t$.
$\epsilon$ is the maximum value of the perturbation corresponding to the maximum allowed SNR.
The parts marked in red relate to the robustness (see Section~\ref{ssec:RoomImpulse}): the Gaussian noise $g\sim\mathcal{N}(0,1)$, the room impulse response $r$ and the pre-appended zero vector $0$.
The parts marked in blue relate to the psychoacoustic masking (see Section~\ref{ssec:Psychoacoustic}): $\alpha$ is a changeable regularization parameter to adjust the focus between model loss and psychoacoustic loss, while $\beta$ is a normalization factor that is selected depending on the length of the audio sample. 
This way, the same $\alpha$ can be used for every sample without worrying about deviating weighting. 
Here, $\text{len}(x)$ is the length of the audio sample, and $\text{len}(x_{\text{ref}})$ is the length of a reference audio sample used for tuning the parameters. 

In the following part, we explain how to choose the $\alpha$ value.
As the best $\alpha$ is different for every audio sample, it is impossible to know the best
value beforehand. 
Thus, we adapt $\alpha$ dynamically during optimization, depending on the success of finding an adversarial example with the current $\alpha$.
We start with a value for $\alpha$ that is small enough to find a successful adversarial example for any audio sample (the noise may be arbitrarily high in this case). 
Then, whenever a successful adversarial example is found, we increase $\alpha$ and the optimization continues with this new value. 
Accordingly, when the optimization does not find a successful adversarial example anymore, $\alpha$ is decreased again. 
This way, the optimization converges to the maximum $\alpha$ that can be used while still being successful. 
This procedure is inspired by \cite{qin2019imperceptible}, though we adapt their strategy to decide when to increase and decrease $\alpha$. 

For our adaptation, we check in every iteration if the transcription of the current perturbation is equal to the target sentence. 
In a counter, we keep track of how many times in a row the perturbation was successful (positive) or unsuccessful (negative). 
Every time the streak breaks, e.g., a correct transcription follows an incorrect transcription, the counter is reset and started in the other direction. 
When it reaches a specific positive or negative value, $\alpha$ is increased or decreased, and the counter is reset. 
To facilitate a gradual growth of $\alpha$, we too set the limit for increasing $\alpha$ lower than the limit for decreasing $\alpha$ (20 vs.\ -100). 
Furthermore, $\alpha$ cannot be decreased below its initial value. 
With this improvement, we try to make the adaptation of $\alpha$ more stable and prevent random factors like the alignment of increase/decrease intervals.


\section{Experiments}
In the following, we compare four methods with respect to their robustness and imperceptibility:
\begin{itemize}
    \item \textbf{Baseline attack}: CW attack with $L_2$ norm adapted for SpeechBrain, 
    \item \textbf{Robust attack}: variation using $L_2$ norm and EOT to train robust adversarial examples,
    \item \textbf{Psychoacoustic attack}: variation using psychoacoustic regularization,
    \item \textbf{Combined attack}: variation using both a psychoacoustic regularization term as well as the EOT algorithm.
\end{itemize}
We outline the experiments in the following subsections, the results are discussed in Section~\ref{sec:Results}.

\paragraph*{Model under attack}
The model we target in our attacks is the \verb|asr-crdnn-rnnlm-librispeech| model, a pretrained SpeechBrain \cite{speechbrain} model\footnote{See \url{https://huggingface.co/speechbrain/asr-crdnn-rnnlm-librispeech}.} and one of the natively supported models of the \verb|robust_speech| framework \cite{olivier2022recent}. 
It is an end-to-end system trained on the full 10 million-word LibriSpeech dataset. 
On the test set, it reported to achieve a WER of 3.09.

\paragraph*{Dataset}
The audio samples we use for our attacks are taken from the LibriSpeech dataset.
LibriSpeech contains about 1000 hours of read English speech, which is segmented and transcribed \cite{panayotov2015librispeech}. 
For our experiments, we use 50 randomly selected audio samples from the test-clean split of the LibriSpeech dataset, which contains relatively clear audio. 
We only select audio files which are transcribed correctly by the model in their original form, apart from minor errors like names.

\paragraph*{Target phrases}
We use four different sentences as the target phrases for our adversarial examples, see Table~\ref{tab:TargetSentences}.
This way, we decrease the risk of accidentally choosing an especially easy or challenging target sentence and distorting the results. 
Additionally, because we choose target phrases of different lengths, we can evaluate how the length of the target phrase affects the success of the adversarial example. 
We used $T_1$ to optimize the hyperparameters, including learning rate and $\alpha$. 
$T_2$, $T_3$ and $T_4$ are chosen from samples in the LibriSpeech dataset. 
None of them belong to the audio samples we attack, and they are chosen only to contain words known to the NN and no names.

\paragraph*{Attack parameters}
We used an initial value of $\alpha = 0.3$, a factor of 1.1 to change it, and a required
success/failure streak of 15/100 to increase/decrease it. 
For the examples using the $L_2$ norm, the learning rate was 0.002. 
For the ones using a psychoacoustic regularization term, we found a learning rate of 0.001 to yield better results. 
All of the examples are trained for at least 5,000 iterations. 

\begin{table}[h]
    \centering
    \caption{Target phrases for the evaluation. 
    $S_1$ was picked for its security relevance, $S_2 - S_4$ are from the LibriSpeech dataset.}
    \begin{tabular}{cL{6cm}}
        $S_1$ & Please open the door \\
        $S_2$ & It is manifest that man is now subject \\
        $S_3$ & But the bear maintained the seat it had taken \\
        $S_4$ & Their assumed character changed with their changing opportunities
    \end{tabular}
    \label{tab:TargetSentences}
\end{table}

\subsection{Exp.~1: simulated environment}
We first test our adversarial examples in a simulated environment.
For this, we evaluate the transcription after transforming the audio file with offset, generated RIR, or Gaussian noise.
Per file, we apply ten different transformations.
For each test, we record the word error rate (WER), and additionally, the signal-to-noise ratio (SNR) of every perturbation.

\subsection{Exp.\ 2: real-world environment}\label{ssec:RealWorld}
\begin{figure}
    \centering
    \includegraphics[width=0.4\textwidth]{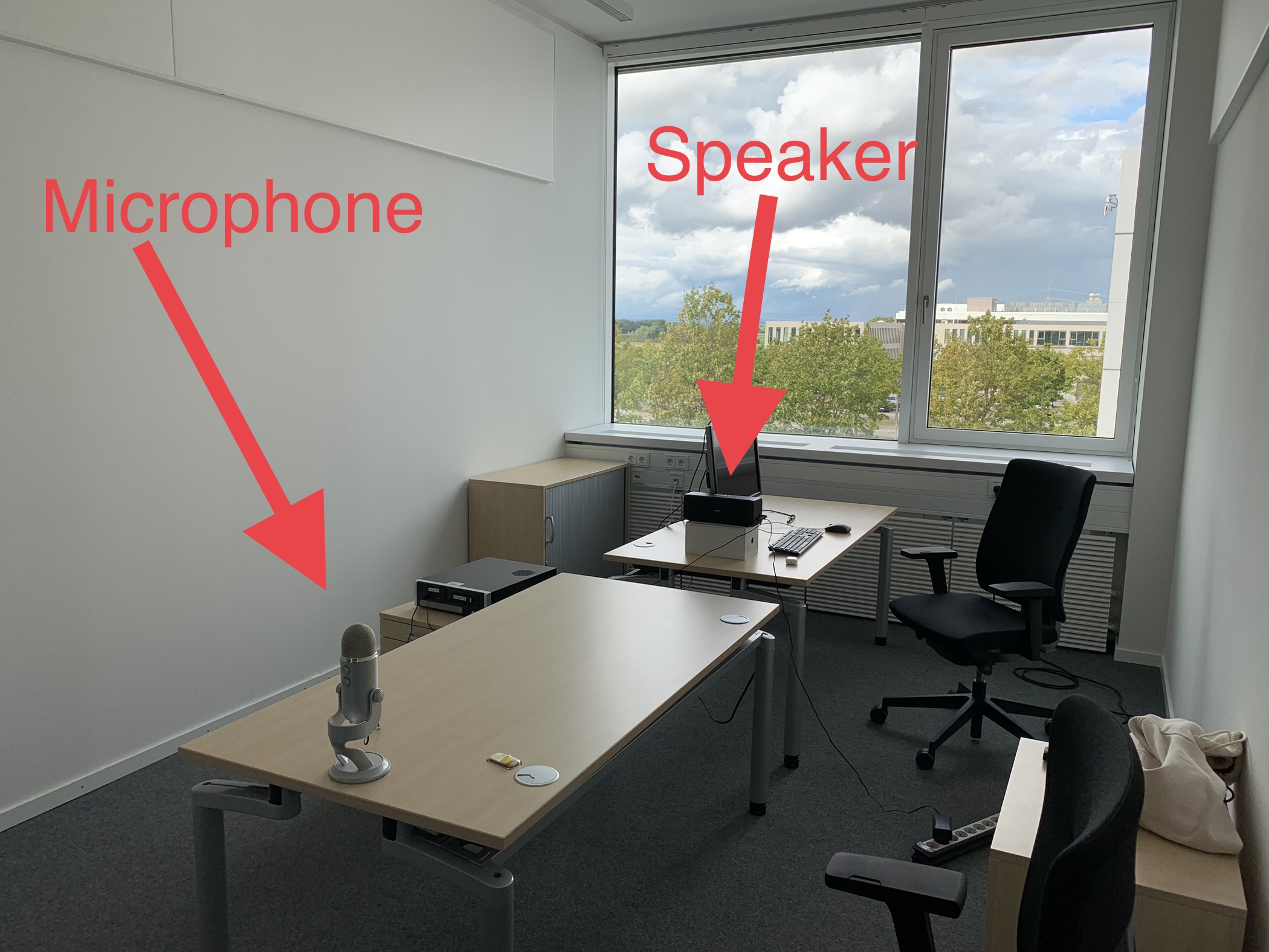}
    \caption{Setup for real-world experiments. 
    Office environment with speaker and microphone at 2~m distance.}
    \label{fig:RobustnessSetup}
\end{figure}

\begin{figure}
    \centering
    \includegraphics[width=0.35\textwidth]{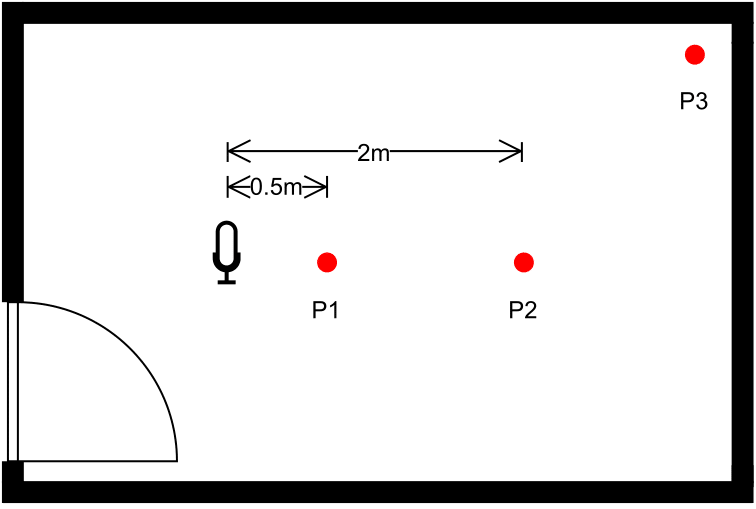}
    \caption{Position of the microphone and the three positions for the speaker that were used to take the measurements.}
    \label{fig:RobustnessPlacement}
\end{figure}

Since the goal for our adversarial examples is to be robust against true over-the-air transmission, we also conduct physical measurements.
For our tests, we use a small office, the layout and furniture in the office can be seen in Figure~\ref{fig:RobustnessSetup}. 
The environment is relatively silent, with occasional background noise. 
With a script, we played pre-generated adversarial examples on the speaker and recorded by the microphone, both connected to the same computer. 
The speaker is a ``DOCKIN D FINE'' speaker using Aux input, and the microphone is a ``Blue -- Yeti'' using an omnidirectional recording pattern. 
Each audio file is played and recorded 10 times to reduce random factors that might influence the transmission. 
The resulting re-recorded audio files are evaluated without applying any additional transformations.
The microphone and speaker are initially placed on a desk in the middle of the room at the same height, at a distance of 50~cm. 
Furthermore, we also test the performance at different positions in the room. 
For this, we conduct tests with the speaker placed at two other locations at 2~m and 3~m distance, see Figure~\ref{fig:RobustnessPlacement}.

Additionally, to test if creating the RIRs dynamically improves performance, we conducted tests with an adapted implementation, where a certain amount of RIRs are generated before optimization and one of them was randomly chosen for each iteration. 
Concretely, we tested with a pool of 32 RIRs and 128 RIRs. 
For each pool, we tested once with the same RIR inputs as the dynamic setup, i.e., a different room for each RIR, and once with RIR inputs from only a single room, i.e., dimensions and reverberation time are identical and only the positions of speaker and listener change.
The attacks generated with these configurations were then evaluated in the simulated environment and over-the-air setup and compared with the results of the dynamically generated RIRs. 
Here, we only generated attacks for target phrase $S_1$.

\subsection{Exp.\ 3: human perceptibility study}
Since it is difficult to accurately evaluate the perceptibility of audio files with technical metrics only, we conducted a human study to obtain definitive results. 
For this study, we recruited 25 people from our colleagues and friends. 
The study was conducted as a browser-based survey, hence no controlled environment. Participants could fill out the survey on their phone or computer at a location of their choice. 
Participants were also allowed to use either speakers or headphones to cover many different setups. 
During the study, each participant was presented several questions. 
Each of the questions contains embedded audio files, which they are then asked to judge on different criteria. 
We used five different question types, each of which is posed several times with different audio samples. 
We ordered the questions in a way to minimize learning bias, placing questions that allow participants to learn what an attack sounds like later in the survey. 
The order of the sub-questions for each question type is randomized for each participant. 
We also asked preliminary information on age, the noise of the listening environment, proficiency in understanding spoken English, and if a headphone or a speaker was used.
Because of how the audio files are embedded in the survey (SoundCloud), a small aesthetic visual representation of the audio file in time format is shown in the embedding. 
We focused on three types of questions: (1) Perceptibility of manipulations, (2) quality ranking, (3) transcription.
We detail the questions in the following.

\paragraph*{Perceptibility of manipulations}
For the first question, the participants are presented with a single audio file and asked to ``decide if the audio is clean, i.e., if there is unusual noise, artifacts, manipulations''.
This question is posed for ten different audio samples, two for each the original, baseline, robust attack, psychoacoustic attack, and combined attack.
In the second question, two audio samples are presented, and the participants are asked to ``decide if there are any differences or if they are identical''. 
The question is asked for eight pairs of audio samples, two for each configuration apart from the robust attack.
The third of the questions presents two audio samples and asks the participants to ``choose which one is the original''. 
The audio samples always contain the original audio and a manipulated version. 
The question is asked for six pairs of audio samples, two for each attack apart from the robust attack.

\paragraph*{Quality ranking}
The participants are presented with four versions of the same audio sample, corresponding to the four different attack configurations. 
The participants are then tasked to ``rank them from best to worst, in terms of quality and clarity''. 
We pose this question for four different audio samples, although each participant is only presented with two of them because the question is quite time intensive.

\paragraph*{Transcription}
Each participant is presented with an audio sample and asked to transcribe the audio. 
They are also explicitly told to note any words they might hear in the background. 
We use five versions of this question, corresponding to the five variants of the audio sample, the original audio and the four different attack configurations. 
Each participant is presented with one of the five versions.

\section{Results}\label{sec:Results}
\subsection{Exp.\ 1: simulated environment}
The results of the tests in the simulated environment can be seen in Table \ref{tab:ExpSimulation}. 
They show that the generation process for adversarial examples fails substantially more often (\%-success) for the two configurations using the EOT algorithm (i.e., robust, combined). 
This indicates that the additional requirement of robustness poses a considerable challenge for the optimizer. 
This stands in contrast to using the psychoacoustic metric, which seems to make the optimization easier when compared to the corresponding variants using the $L_2$ norm as regularization term (i.e., base vs.\ psy.\ and rob.\ vs.\ comb.). 
It also shows that every variant has a harder time generating adversarial examples the longer the target phrase is, e.g., $S_1$ to $S_4$.

The base and psychoacoustic method often succeed to find an adversarial example during optimization. 
However, these adversarial examples never successfully reach the target phrase in the simulated environment containing reverberation and other transformations, as those are not considered during the generation process, thus leading to a WER of 100. 
While the algorithms considering robustness achieve a lower success rate in generating adversarial examples, their overall WER is also lower, thus better.
This shows 
that using the integrated algorithm, the adversarial examples can be made robust against arbitrary RIRs and achieve low WER in a simulated environment, especially for shorter sentences. 
As Table \ref{tab:ExpSimulation} shows, the WER for the combined approach is lower than the WER for the robust variant, further indicating that the psychoacoustic metric makes optimization less challenging compared to the $L_2$ norm.

\begin{table}[h]
    \setlength{\tabcolsep}{4pt}
    \centering
    \begin{tabular}{ccrrrrr}
        \toprule
        & \textbf{Target} & \textbf{$S_1$} & \textbf{$S_2$} & \textbf{$S_3$} & \textbf{$S_4$} & \textbf{Mean} \\
        \midrule
        \multirow{3}{*}{\rotatebox[origin=c]{90}{base}} & WER & 100.00 & 100.00 & 100.00 & 100.00 & 100.00 \\
        & SNR & 24.38 & 23.67 & 21.63 & 21.00 & 22.74 \\
        & Success & 100.0\% & 98.0\% & 96.0\% & 84.0\% & 94.5\% \\
        \midrule
        \multirow{3}{*}{\rotatebox[origin=c]{90}{rob.}} & WER & 12.46 & 50.04 & 23.20 & 60.44 & 36.54 \\
        & SNR & 14.76 & 17.52 & 15.47 & 15.67 & 15.64 \\
        & Success & 98.0\% & 54.0\% & 86.0\% & 48.0\% & 71.5\% \\
        \midrule
        \multirow{3}{*}{\rotatebox[origin=c]{90}{psy.}} & WER & 100.00 & 100.00 & 100.00 & 100.00 & 100.00 \\
        & SNR & 7.90 & 8.52 & 6.84 & 4.74 & 7.06 \\
        & Success & 100.0\% & 98.0\% & 98.0\% & 86.0\% & 95.5\% \\
        \midrule
        \multirow{3}{*}{\rotatebox[origin=c]{90}{comb.}} & WER & 2.85 & 27.94 & 20.00 & 54.98 & 26.44 \\
        & SNR & 5.36 & 2.30 & 3.32 & 2.48 & 3.60 \\
        & Success & 100.0\% & 74.0\% & 82.0\% & 50.0\% & 76.5\% \\   
        \bottomrule
    \end{tabular}
    \caption{The WER, SNR, and percentage of successfully generated examples for the four used attack variants in a simulated environment. The percentage of successfully found examples is the percentage of attacked audio samples, where during at least one iteration of the optimization process, the current  perturbation produced the correct transcription.}
    \label{tab:ExpSimulation}
\end{table}

A high SNR usually signifies that the generated example is better and was easier to generate than an attack with a lower SNR. 
However, this only works for the baseline and robust attack, as both other attacks do not optimize the SNR but rather a different psychoacoustic metric.
Although only subtle, the average SNR of generated examples steadily declines from $S_1$ to $S_4$ with more challenging targets.
However, between individual target samples, there is an extensive range of different SNRs spread relatively evenly. 
For example, in the case of the baseline model with target phrase $S_1$, the values range from about 17 to 33.
This could indicate that the difficulty of an attack depends on the target phrase, similar to the discussion in \cite{carlini2018audio}.

\subsection{Exp.~2: real-world environment}
For our over-the-air measurements, we used the setup described in Section \ref{ssec:RealWorld}. 
To compare the performance of all four configurations, we took the measurements of every configuration and every target phrase.
Table~\ref{tab:ExpRealWorldP1} displays the results for position P1 at a distance of 50cm, see Figure~\ref{fig:RobustnessPlacement}.
\begin{table}[h]
    \centering
    \begin{tabular}{crrrr}
        \toprule
         & \textbf{base.} & \textbf{rob.} & \textbf{psy.} & \textbf{comb.} \\
         \midrule
        WER & 98.96 & 99.95 & 94.80 & 97.40 \\
        \bottomrule
    \end{tabular}
    \caption{Average word error rate (WER) for the different attacks in the real-world environment.
    Even though trained with room impluse responses, the robust attack exhibits the worst WER.}
    \label{tab:ExpRealWorldP1}
\end{table}
As seen in Table~\ref{tab:ExpRealWorldP1}, all methods suffer dramatic decreases in performance compared to the tests in the simulation from Table~\ref{tab:ExpSimulation}. 
This failure is expected with the two configurations not optimized for robustness\footnote{Although both of these configurations have a WER of less than 100, when investigating the underlying results, it is discovered that the WER below 100 is not because of performance. Instead, it does not change the target audio at all, and there are just some original transcriptions that are close enough to the target phrase for a WER below 100.}. 
Both methods using RIRs during training perform slightly better, although with WERs of 94.80 and 97.40 respectively, they only rarely come close to the target transcription. 
However, when inspecting the individual transcriptions, there is a noticeable improvement to the methods not using the EOT algorithm in that singular words from the target phrase are frequently transcribed correctly. 
However, none of the attacks ever reach the exact target phrase.

To evaluate how robust our attacks are from different positions in the room, we additionally perform measurements from the positions in Figure~\ref{fig:RobustnessPlacement}. 
With this, we evaluate the robust attack.
\begin{table}[h]
    \centering
    \begin{tabular}{crrrrr}
        \toprule
        \textbf{Position} & \textbf{$S_1$} & \textbf{$S_2$} & \textbf{$S_3$} & \textbf{$S_4$} & \textbf{Mean} \\
        \midrule
        P1 & 95.03 & 95.97 & 91.91 & 98.23 & 94.80 \\
        P2 & 99.84 & 97.78 & 93.95 & 99.69 & 97.64 \\
        P3 & 99.74 & 98.94 & 97.96 & 99.69 & 99.04 \\
        \bottomrule
    \end{tabular}
    \caption{WER for different microphone/speaker setups in the real-world environment, see Figure~\ref{fig:RobustnessPlacement}.
    The greater the distance, the higher the WER.}
    \label{tab:ExpRealWorldPositions}
\end{table}

As seen in Table~\ref{tab:ExpRealWorldPositions}, the effectiveness quickly decreases with increasing distance (P2 has 2m, P3 has 3m). 
This is the case for all sentences. 
Moreover, for all but $S_3$, the effectiveness decreases to levels of examples without robustness training.
Generally, there appears to be a correlation between the distance and placement of the microphone and speaker, despite the success in a simulated environment at different positions.
While we achieve some success with certain configurations, overall, our attacks show only minimal success in over-the-air scenarios. 
Especially when comparing the results to those in the simulated environment, which shows that the algorithm generates adversarial examples that are successful most of the time when transforming them with the generated RIRs. 
This discrepancy could indicate that the simulated environment used for the transformations is not a close enough approximation of the physical conditions and that relevant transformations might be missing. 
A candidate for this could be the changes applied to the audio by the microphone and speaker. 
Different speakers and microphones have different sound characteristics, influencing the audio played and recorded by them.

Furthermore, we evaluated if the results improve when the adversarial examples are trained on rooms more similar to the one in which they are tested. 
To do this, we adjusted the inputs to the RIR generator to more closely resemble the actual room we used for our experiments. 
We only changed the interval the reverberation time is chosen from. 
The entire possible interval is [0.2, 0.8]. 
Because we do not know the real reverberation time of the room, we split the interval and conduct tests with both parts.
First, we conducted experiments with a higher and a lower distribution. 
The better performing of these distributions was then split further to identify the interval leading to the best performance. 
We only used $S_1$ as the target phrase. 
The results can be found in Table~\ref{tab:ExpRealWorldIntervals}.

\begin{table*}[h]
    \centering
    \begin{tabular}{crrrrr}
        \toprule
        \textbf{Rev.\ interval} & \textbf{$[0.2, \, 0.5]$} & \textbf{$[0.4, \, 0.8]$} & \textbf{$[0.2, \, 0.3]$} & \textbf{$[0.3, \, 0.4]$} & \textbf{$[0.4, \, 0.5]$} \\
        \midrule
        WER & 81.39 & 95.03 & 86.65 & 77.97 & 74.82 \\ 
        \bottomrule
    \end{tabular}
    \caption{Comparison of the WER for different reveberation intervals used for the robustness attack.
    Most likely, the reveberation time of the real-world room setup was in the interval $[0.4, \, 0.5]$.}
    \label{tab:ExpRealWorldIntervals}
\end{table*}

From the first two chosen intervals, it is apparent that $[0.2, \, 0.5]$ has a substantially lower WER, indicating that the real RIR lies in that interval. 
When splitting the interval further, we can observe a lowering WER, which culminates in the interval $[0.4, \, 0.5]$.
This leads us to believe the real reverberation time of the room we experiment in is in this interval. 
Although both of the larger intervals contain the interval $[0.4, \, 0.5]$, the interval $[0.2, \, 0.5]$ performs much better than $[0.4, \, 0.8]$.
While none of the other attacks achieved entirely correct target transcriptions, the attacks using the interval $[0.4, \, 0.5]$ achieved a success rate of 8.40\%, even achieving a success rate of 100\% for 2 out of the 50 used audio files.

Some intriguing differences arise when comparing our results to the results of \cite{schoenherr-19-psychoacoustic}. 
Although \cite{schoenherr-19-psychoacoustic} conduct most of their over-the-air measurements in a lab room, which is considerably larger than the office we use, they also conduct a series of measurements comparing performance over different room types, where they show that adversarial examples perform better in larger rooms. 
One of the rooms they use for this measurement series is an office, which is still larger than ours but comparable.
In this office, they achieved a WER of 74.0, which is about the same as the WER of our attacks with adapted reverberation time (74.82), but better than that of our attacks with a large reverberation time interval. 
However, contrary to our results, which show that an adapted reverberation time interval leads to a smaller WER, their results show the opposite relationship, where a reverberation time more closely adapted to the room increases the WER.
However, while such comparisons between attacks can be interesting, it is crucial to remember that they are only comparable to a minimal extent, as immense differences exist between the used methods. 
\cite{schoenherr-19-psychoacoustic} use a different framework (Kaldi), model, dataset, microphone, speaker, and room with a different size and furnishings. 
This leaves only similarities on a very fundamental level.

In the last part of our over-the-air study, we considered pre-generated room impulse responses with dynamically generated ones.
Both methods with 32 and 128 RIRs have a substantially higher WER than the normal robust attack. 
However, the configuration with 128 RIRs substantially improved from the one with 32, almost halving the WER. 
Interestingly, there is only a minimal difference between RIRs generated with only a single room and 32/128 different rooms.
This seems to indicate that the relative positioning of the speaker and the listener might be substantially more influential than the room’s specifications. Still, the methods with different rooms perform slightly better than those without.
In the over-the-air measurements, the pattern is repeated, although with substantially worse WER.

\begin{table}[h]
    \centering
    \begin{tabular}{crrrr}
        \toprule
         & \multicolumn{2}{c}{\textbf{Simulation}} & \multicolumn{2}{c}{\textbf{Over-the-air}} \\
        \textbf{RIR generation} & \textbf{WER} & \textbf{Correct} & \textbf{WER} & \textbf{Correct} \\
        \midrule
        dynamic & 10.67 & 86.16\% & 94.80 & 0.00\% \\
        32 (one) & 87.15 & 14.00\% & 100.00 & 0.00\% \\
        32 (various) & 82.41 & 18.36\% & 100.00 & 0.00\% \\
        128 (one) & 48.62 & 47.84\% & 99.85 & 0.00\% \\
        128 (various) & 47.80 & 45.71\% & 99.55 & 0.00\% \\
        \bottomrule
    \end{tabular}
    \caption{Performance comparison of the robust attack using dynamically generated RIRs or a fixed pool of RIRs, either with a random room for each RIR or the same room for all RIRs. 
    Only successfully generated attacks are considered.}
    \label{tab:RIRComparison}
\end{table}

\subsection{Exp.~3: human perceptibility study}
We collected responses from 25 participants. 
Most of the 25 them are young adults, with 72\% of participants between 21 and 30 years old and 20\% between 31 and 40 years old. 
This predominately low age of participants is convenient for our study, as younger people usually have better hearing, which allows for better estimation of the upper bound of perceptibility. 
With 64\%, about two-thirds of the listeners used headphones to answer the survey and the rest used speakers. 
Most participants were in a quiet environment, with 88\% describing their environment as either ``completely silent'' or ``mostly silent''. 
The rest described their environment as only ``a little noisy''. 
Out of all participants, 92\% rated their ability to understand spoken English as either ``good'' or ``very good'', with the remaining 8\% rating it as ``ok''.

\begin{table}[h]
    \centering
    \begin{tabular}{crrrrr}
        \toprule
         & \textbf{original} & \textbf{base.} & \textbf{rob.} & \textbf{psy.} & \textbf{comb.} \\
        \midrule
        \% clean & 82\% & 14\% & 2\% & 36\% & 6\% \\
        identical\ to\ orig. & 90\% & 0\% & - & 44\% & 0\% \\
        identify orig. & - & 80\% & - & 80\% & 86\% \\
        quality\ rank. & - & 2.58 & 3.68 & 1.18 & 2.56 \\
        \bottomrule
    \end{tabular}
    \caption{Results from the listener experiments: (1) Is this clean audio? (2) Is this identical to the original? (3) Is this the original? (4) Quality ranking.}
    \label{tab:ExpHuman}
\end{table}

Our results in Table~\ref{tab:ExpHuman} show that the psychoacoustic attack is much harder to detect for humans than any attack based on the $L_2$ norm.
Further, we find that the combined attack is rated to be of equal quality to the baseline attack.
Listeners evaluate the original files as 82\% clean, out of the adversarial examples generated with respect to psychoacoustic, still 36\% are rated clean.
Similarly, original files are reliably identified as identical to the original, for psychoacoustic samples, still 44\% stand this comparison.
For any attack, at least 80\% correctly identified the original file.
While the psychoacoustic attack has the highest quality ranking, the baseline and the combined attack perform very similarly.

Lastly, to verify that the adversarial examples do not change the human transcription of the target sentence, we asked participants to transcribe the audio they hear in different versions of the attack. 
Each participant was only presented with one of the versions.
Every version was transcribed correctly all of the time. 
Additionally, only three participants heard any other words in the background. 
Two of the three notice additional words in the combined attack but were not able to (correctly) identify them, one of them describing it as a kind of echo. 
The third one made out additional words in the baseline and could correctly identify one part of the target phrase $S_1$ (``please'').
Therefore, none of the attacks inhibit the ability of humans to transcribe the audio correctly, and in the vast majority of cases, none of the words in the target phrase can be heard. 
Especially, none are correctly identified for the psychoacoustic attacks.

\subsection{Summary}
Overall, our psychoacoustic metric provides a clear improvement over the $L_2$ norm in terms of perceptibility and perceived quality: both the psychoacoustic attack and the combined attack score better than the baseline and the robust attack, respectively. 
The psychoacoustic attack achieves good scores in imperceptibility, with 44\% of participants unable to differentiate between the original and the adversarial example. 
However, it is still not close to the desired values of the original audio. 
Additionally, while the combined attack ranks higher than the robust attack in terms of quality, it is only even with the baseline while being far less successful. 
Furthermore, it achieves no success in the questions focusing on imperceptibility. 
Nonetheless, all attacks are still correctly transcribed by humans, and the target phrase is not heard.

\section{Conclusion and outlook}

In this work, we analysed how useful an integrated algorithm is for creating audio adversarial examples that should be both robust and imperceptible. 
We showed that dynamically creating RIRs for generating robust audio adversarial examples is a viable method, generating more successful attacks than a fixed dataset of RIRs, especially if this dataset is small.
We supported these results with over-the-air measurements, which show that the generated adversarial examples experience at least limited success and that this improvement with dynamically generated RIRs is reflected in physical conditions.
However, the overall results indicate that is challenging for the optimizer to consider robustness and adversarial success in just one integrated step.
While we have used only one RIR model, it is left for future work to consider and compare other robustness models.

Furthermore, we refined the psychoacoustic model implementation to achieve imperceptible adversarial examples and evaluate its effectiveness in a human perceptibility study. 
While a substantial number of participants could not perceive a difference between the original audio and the adversarial example, a substantial part of them could still hear differences.
Finally, we show how an attacker might individualize the training parameters according to the target room to obtain better over-the-air results.
In future work, other models for creating RIRs might be included in an integrated approach and compared to existing works like \cite{qin2019imperceptible, schoenherr-19-psychoacoustic} with respect to robustness, perceptibility and runtime.


\bibliographystyle{IEEEtran}
\bibliography{mybib}

\begin{thebibliography}{10}
\providecommand{\url}[1]{#1}
\csname url@samestyle\endcsname
\providecommand{\newblock}{\relax}
\providecommand{\bibinfo}[2]{#2}
\providecommand{\BIBentrySTDinterwordspacing}{\spaceskip=0pt\relax}
\providecommand{\BIBentryALTinterwordstretchfactor}{4}
\providecommand{\BIBentryALTinterwordspacing}{\spaceskip=\fontdimen2\font plus
\BIBentryALTinterwordstretchfactor\fontdimen3\font minus
  \fontdimen4\font\relax}
\providecommand{\BIBforeignlanguage}[2]{{%
\expandafter\ifx\csname l@#1\endcsname\relax
\typeout{** WARNING: IEEEtran.bst: No hyphenation pattern has been}%
\typeout{** loaded for the language `#1'. Using the pattern for}%
\typeout{** the default language instead.}%
\else
\language=\csname l@#1\endcsname
\fi
#2}}
\providecommand{\BIBdecl}{\relax}
\BIBdecl

\bibitem{abdullah2021sok}
H.~Abdullah, K.~Warren, V.~Bindschaedler, N.~Papernot, and P.~Traynor, ``Sok:
  The faults in our asrs: An overview of attacks against automatic speech
  recognition and speaker identification systems,'' in \emph{2021 IEEE
  symposium on security and privacy (SP)}.\hskip 1em plus 0.5em minus
  0.4em\relax IEEE, 2021, pp. 730--747.

\bibitem{carlini2018audio}
N.~Carlini and D.~Wagner, ``Audio adversarial examples: Targeted attacks on
  speech-to-text,'' in \emph{2018 IEEE Security and Privacy Workshops (SPW)},
  2018, pp. 1--7.

\bibitem{qin2019imperceptible}
Y.~Qin, N.~Carlini, G.~Cottrell, I.~Goodfellow, and C.~Raffel, ``Imperceptible,
  robust, and targeted adversarial examples for automatic speech recognition,''
  in \emph{International conference on machine learning}.\hskip 1em plus 0.5em
  minus 0.4em\relax PMLR, 2019, pp. 5231--5240.

\bibitem{athalye2018synthesizing}
A.~Athalye, L.~Engstrom, A.~Ilyas, and K.~Kwok, ``Synthesizing robust
  adversarial examples,'' in \emph{International conference on machine
  learning}.\hskip 1em plus 0.5em minus 0.4em\relax PMLR, 2018, pp. 284--293.

\bibitem{schonherr2020imperio}
\BIBentryALTinterwordspacing
L.~Sch\"{o}nherr, T.~Eisenhofer, S.~Zeiler, T.~Holz, and D.~Kolossa, ``Imperio:
  Robust over-the-air adversarial examples for automatic speech recognition
  systems,'' in \emph{Annual Computer Security Applications Conference}, ser.
  ACSAC '20.\hskip 1em plus 0.5em minus 0.4em\relax New York, NY, USA:
  Association for Computing Machinery, 2020, p. 843–855. [Online]. Available:
  \url{https://doi.org/10.1145/3427228.3427276}
\BIBentrySTDinterwordspacing

\bibitem{dorr2020towards}
T.~D{\"o}rr, K.~Markert, N.~M. M{\"u}ller, and K.~B{\"o}ttinger, ``Towards
  resistant audio adversarial examples,'' in \emph{Proceedings of the 1st ACM
  Workshop on Security and Privacy on Artificial Intelligence}, 2020, pp.
  3--10.

\bibitem{ijcai2019p741}
\BIBentryALTinterwordspacing
H.~Yakura and J.~Sakuma, ``Robust audio adversarial example for a physical
  attack,'' in \emph{Proceedings of the Twenty-Eighth International Joint
  Conference on Artificial Intelligence, {IJCAI-19}}.\hskip 1em plus 0.5em
  minus 0.4em\relax International Joint Conferences on Artificial Intelligence
  Organization, 7 2019, pp. 5334--5341. [Online]. Available:
  \url{https://doi.org/10.24963/ijcai.2019/741}
\BIBentrySTDinterwordspacing

\bibitem{chen2020metamorph}
T.~Chen, L.~Shangguan, Z.~Li, and K.~Jamieson, ``Metamorph: Injecting inaudible
  commands into over-the-air voice controlled systems,'' in \emph{Network and
  Distributed Systems Security (NDSS) Symposium}, 2020.

\bibitem{olivier2022recent}
R.~Olivier and B.~Raj, ``Recent improvements of asr models in the face of
  adversarial attacks,'' \emph{arXiv preprint arXiv:2203.16536}, 2022.

\bibitem{speechbrain}
M.~Ravanelli, T.~Parcollet, P.~Plantinga, A.~Rouhe, S.~Cornell, L.~Lugosch,
  C.~Subakan, N.~Dawalatabad, A.~Heba, J.~Zhong, J.-C. Chou, S.-L. Yeh, S.-W.
  Fu, C.-F. Liao, E.~Rastorgueva, F.~Grondin, W.~Aris, H.~Na, Y.~Gao, R.~D.
  Mori, and Y.~Bengio, ``{SpeechBrain}: A general-purpose speech toolkit,''
  2021, arXiv:2106.04624.

\bibitem{chen2019shapeshifter}
S.-T. Chen, C.~Cornelius, J.~Martin, and D.~H. Chau, ``Shapeshifter: Robust
  physical adversarial attack on faster r-cnn object detector,'' in
  \emph{Machine Learning and Knowledge Discovery in Databases: European
  Conference, ECML PKDD 2018, Dublin, Ireland, September 10--14, 2018,
  Proceedings, Part I 18}.\hskip 1em plus 0.5em minus 0.4em\relax Springer,
  2019, pp. 52--68.

\bibitem{10.1145/3560830.3563733}
\BIBentryALTinterwordspacing
P.~A. Sava, J.-P. Schulze, P.~Sperl, and K.~B\"{o}ttinger, ``Assessing the
  impact of transformations on physical adversarial attacks,'' in
  \emph{Proceedings of the 15th ACM Workshop on Artificial Intelligence and
  Security}, ser. AISec'22.\hskip 1em plus 0.5em minus 0.4em\relax New York,
  NY, USA: Association for Computing Machinery, 2022, p. 79–90. [Online].
  Available: \url{https://doi.org/10.1145/3560830.3563733}
\BIBentrySTDinterwordspacing

\bibitem{iso1993information}
{ISO/IEC 11172-3}, ``Information technology -- coding of moving pictures and
  associated audio for digital storage media at up to about 1.5 mbit/s -- part
  3: Audio,'' 1993.

\bibitem{lin2015audio}
Y.~Lin, W.~H. Abdulla \emph{et~al.}, \emph{Audio watermark}.\hskip 1em plus
  0.5em minus 0.4em\relax Springer, 2015, vol. 146.

\bibitem{schoenherr-19-psychoacoustic}
L.~Sch{\"o}nherr, K.~Kohls, S.~Zeiler, T.~Holz, and D.~Kolossa, ``Adversarial
  attacks against automatic speech recognition systems via psychoacoustic
  hiding,'' in \emph{Symposium on Network and Distributed System Security
  (NDSS)}, 2019.

\bibitem{9747846}
A.~Ratnarajah, S.-X. Zhang, M.~Yu, Z.~Tang, D.~Manocha, and D.~Yu, ``Fast-rir:
  Fast neural diffuse room impulse response generator,'' in \emph{ICASSP 2022 -
  2022 IEEE International Conference on Acoustics, Speech and Signal Processing
  (ICASSP)}, 2022, pp. 571--575.

\bibitem{panayotov2015librispeech}
V.~Panayotov, G.~Chen, D.~Povey, and S.~Khudanpur, ``Librispeech: an asr corpus
  based on public domain audio books,'' in \emph{2015 IEEE international
  conference on acoustics, speech and signal processing (ICASSP)}.\hskip 1em
  plus 0.5em minus 0.4em\relax IEEE, 2015, pp. 5206--5210.

\end{thebibliography}

\end{document}